\pdfoutput=1
\documentclass[sagev,times]{sagej}

\ifdefined\pdfsuppresswarningpagegroup
\fi

\usepackage[T1]{fontenc}
\usepackage[utf8]{inputenc}
\usepackage{microtype}
\usepackage[hidelinks]{hyperref}
\usepackage{siunitx}

\title{Compression Hurts, Pooling Helps: Information Loss in Rayleigh-Scale Estimation from B-Mode Ultrasound}
\runninghead{Compression Hurts, Pooling Helps}
\author{Hudson Smith\affilnum{1} and Ahmer Raza\affilnum{1}}
\affiliation{\affilnum{1} Clemson University}
\corrauth{Hudson Smith}
\email{dane2@clemson.edu; araza@clemson.edu}
\def\journalname{Preprint}

\usepackage{etoolbox}
\makeatletter
\def\preprint@gobble#1\end{\end}
\patchcmd{\@maketitle}{\journalname\\}{\journalname\preprint@gobble}{}{\errmessage{sagej's title box changed; update the preprint patch}}
\patchcmd{\ps@sagepage}{\ \volumenumber(\issuenumber)}{}{}{\errmessage{sagej's running head changed; update the preprint patch}}
\makeatother

\newcommand{\FisherMinInflation}{34.98}
\newcommand{\FisherMinInflationPrecise}{34.9804}
\newcommand{\FisherMinSigma}{3.694}
\newcommand{\FisherBestCaseWindows}{340}

\begin{abstract}

  Clinical B-mode images are widely available as potential data sources for quantitative ultrasound (QUS) analysis for tissue characterization. However, standard clinical ultrasound devices apply unknown log-compression to RF envelope data before display and storage. Previous work has demonstrated estimation of the underlying RF envelope statistics in the presence of an unknown compression law. Using Fisher information analysis, we show that finite-offset log compression causes severe information loss when estimating the Rayleigh scale $\sigma$, which controls diffuse speckle. For a single image window, unknown compression raises the minimum achievable variance for unbiased estimation of $\sigma$ by a compression-independent factor of approximately $\FisherMinInflation$. When $M$ equal-sized windows share the same unknown compression settings, the excess variance decays as $1/M$; even in the most favorable regime, reducing the variance inflation factor below $1.1$ requires $\FisherBestCaseWindows$ windows. Our analysis treats the contrast parameter $a$ as unknown and the boundary offset $b$ as known; estimating $b$ experimentally shows even larger variance. We validate this theory using synthetic estimation experiments and demonstrate RF-scale recovery on real RF-envelope windows from the OASBUD dataset. Together, these results clarify the limitations of using routine B-mode images for QUS.
\end{abstract}

\begin{document}

\maketitle

\section{Introduction}
The low cost, speed, and noninvasive nature of B-mode ultrasound (US) have established it as a primary imaging modality across emergency medicine, cardiology, obstetrics, and oncology. As a result, medical centers have collected large archives of B-mode scans which could be repurposed for quantitative ultrasound (QUS) applications such as characterizing attenuation rates, scatterer densities, and other tissue properties. Unfortunately, stored B-mode data are optimized for visual interpretation by trained clinicians, not for QUS applications. Beamformed echo signals undergo a complex display pipeline, including envelope detection, depth compensation, log-compression, scan conversion, quantization, and more. The exact sequence of steps depends on vendor-specific design and device settings. Recent reproducibility studies show that scanner settings, beamforming method, and display dynamic range can affect quantitative features extracted from B-mode images \cite{li2022reproducibility,seoni2023texture}. Moreover, vendor documentation usually lacks key details about the transforms, and system settings are not stored alongside the scan data. For these reasons, the QUS community has preferred raw channel-level or RF envelope data sources over these B-mode images.

The Rayleigh distribution is a classical distributional description of RF signal envelopes for fully developed speckle. When many weak, randomly phased scatterers contribute to a resolution cell and no coherent component dominates, the real and imaginary components of the backscattered field are approximately Gaussian by a central-limit argument. The detected envelope is then Rayleigh distributed and the intensity is exponential \cite{goodman1975speckle,burckhardt1978speckle,wagner1987statistical}. In this regime, the Rayleigh scale can serve as a simple proxy for local diffuse echo strength after the effects of system gain, attenuation, focusing, and display processing are accounted for. Departures from the fully developed speckle setting motivate richer models, including Rice, Nakagami, K, and homodyned-K families, which account for coherent components, scatterer clustering, or departures from the classical central-limit regime \cite{destrempes2025theoretical}. In addition, the validity and interpretation of distributional parameters also depend on system linearity, scatterer geometry, sampling conditions, and estimator uncertainty \cite{christensen2024systematized}. Despite these complications, the Rayleigh case remains a useful baseline for asking whether B-mode display compression preserves even the simplest envelope-scale information.

This raises the question, what information may be recovered from B-mode images? Many studies use displayed ultrasound images directly, without attempting to undo the display pipeline. Radiomics workflows extract shape, histogram, and texture features from display-processed US images and then train predictive models for diagnosis, quality grading, risk stratification, or treatment response \cite{cloutier2021primer,jia2022radiomics}. More recent deep-learning studies follow the same logic with learned features rather than hand-crafted features; for example, liver-steatosis models have been trained from ordinary ultrasound images when raw RF data are unavailable \cite{li2021steatosisdl}. These results align with the purpose of B-mode display processing itself, namely to convert raw echo data into images that preserve clinically useful visual information for diagnosis. At the same time, these studies usually target a clinical biomarker, not calibrated acoustic parameters describing the underlying tissues.

More relevant to the current study, an older line of work has treated the display pipeline itself as part of the estimation problem. Early work characterized scanner signal processing and showed that nonlinear amplification and log compression could be compensated in some settings \cite{crawford1993compensation,kaplan1994logcompressed,dutt1996logcompressed}. Subsequent methods used statistical envelope models to decompress B-mode images, estimate scanner compression parameters, or recover envelope-like images from displayed grayscale data \cite{prager2003decompression,sanches2003compensation,vegas2011realistic}. Related work reached a similar conclusion in the context of video data: inverse display mappings can recover useful echo-power information under controlled conditions \cite{payen2013echopower}. Together, these studies frame QUS from B-mode as a limited but meaningful estimation problem. However, the precise limits of this problem have not been quantified, and the practical implications for QUS remain unclear.

The current work builds directly on the Seabra--Sanches approach to RF-envelope recovery from B-mode images \cite{seabra2008modeling,seabra2012rf}. Seabra and Sanches model the displayed image as a Rayleigh-distributed speckle amplitude that has been compressed through a log transformation. Within homogeneous image regions, they exploit the distributional properties of log-Rayleigh speckle to estimate the global contrast and brightness parameters and invert the fitted display mapping. Their results demonstrate that B-mode images retain sufficient statistical structure to support recovery of the underlying Rayleigh scale. Our study provides an information-theoretic account of this recovery problem. We quantify how uncertainty in the compression contrast limits Rayleigh-scale estimation, and determine how pooling windows with shared compression settings mitigates this loss.

Our primary contributions are:
\begin{enumerate}
    \item We show that only a scanner-normalized Rayleigh scale can be identified from log-compressed B-mode data, not the absolute scale.
    \item We show that Rayleigh scale estimation in the presence of unknown log-compression inflates the best unbiased estimator variance by at least $34.98$ relative to the known-compression case; that is, relative to estimating the Rayleigh scale directly from RF envelope data.
    \item We show that pooling $M$ display windows with shared compression settings shrinks the variance inflation penalty by a factor proportional to $1/M$, partially mitigating the negative consequences of unknown compression.
    \item We validate these theoretical results using extensive simulations, and test their practical applicability using real-world RF-envelope data from OASBUD.
\end{enumerate}

\section{Statistical Model}

\subsection{Log-compressed Rayleigh envelope model}
Let $Y^*$ be the unobserved echo envelope at a given pixel and $Z$ be the observed B-mode pixel measurement. We treat each analysis window as locally homogeneous with fully developed speckle, so that the envelope samples are Rayleigh distributed with a common scale parameter. Under a finite-offset log-compression rule, the resulting observation model is
\begin{subequations}
  \label{eq:model_original}
  \begin{align}
    Y^* &\sim \mathrm{Rayleigh}(\sigma^*),
    \label{eq:model_original_rayleigh}
    \\
    Z &= a \log(bY^*+c) + d,
    \label{eq:model_original_observation}
  \end{align}
\end{subequations}
where $\sigma^*$ is the physical Rayleigh scale parameter controlling the strength of the diffuse speckle, and $(a,b,c,d)$ are unknown nuisance parameters controlling the log-compression function. In the large-$\sigma^*$ limit, this model reduces to the conventional log-compression formula $Z=20\log_{10}(Y^*/Y_\mathrm{max})$ with $a \sim 20/\log(10) \approx 8.686$ and $Y_\mathrm{max} \approx \exp(-d/a)/b$.

\subsection{Identifiability and scanner-normalized scale}
Some of the parameters in Eq.~\eqref{eq:model_original} are strictly non-identifiable. To isolate the parameters, write $Y^*=\sigma^*X$ with $X\sim \mathrm{Rayleigh}(1)$ and factor out $c$ inside the logarithm. This leaves only the combinations
\begin{align}
    \sigma &\equiv \frac{b\sigma^*}{c},
    \label{eq:identified-parameters-sigma}
    \\
    b' &\equiv a\log c+d.
    \notag
\end{align}
Thus, the observed distribution depends on the physical scale $\sigma^*$ only through the scanner-normalized scale $\sigma$, and on $(c,d)$ only through the brightness offset $b'$. The ambiguity in $b'$ only affects nuisance parameters. By contrast, the ambiguity in $\sigma$ means that, when $b/c$ is unknown, the true Rayleigh scale $\sigma^*$ is identifiable only up to an unknown multiplicative factor. In pooled settings where $(b,c)$ are shared across windows, this still permits identification of relative Rayleigh scales across windows, even though the absolute scale remains unidentified.

After renaming $b'$ as $b$, the model becomes
\begin{subequations}
  \label{eq:model_ss}
  \begin{align}
    X &\sim \mathrm{Rayleigh}(1),
    \label{eq:model_ss_rayleigh}
    \\
    Z &= a \log(1+\sigma X) + b.
    \label{eq:model_ss_observation}
  \end{align}
\end{subequations}
This is equivalent to the model form used by Seabra and Sanches \cite{seabra2012rf}.
The remaining parameters $(\sigma, a, b)$ are technically identifiable from the observed data, but $\sigma$ is the identifiable scanner-normalized Rayleigh scale rather than the absolute physical scale $\sigma^*$.

\subsection{Observed density and likelihood}

The inverse image of an observation on the unit-Rayleigh scale is
\begin{equation}
  x(z;\sigma,a,b)=\frac{\exp\!\left((z-b)/a\right)-1}{\sigma}.
  \label{eq:paper-inverse-image}
\end{equation}
The observed density follows by applying the change of variables from $X$ to $Z$. Since $dx/dz=(1+\sigma x(z))/(a\sigma)$, we have
\begin{equation}
  f_Z(z\mid \sigma,a,b)
  =
  \frac{x(z)(1+\sigma x(z))}{a\sigma}
  \exp\!\left(-\frac{x(z)^2}{2}\right)
  \mathbf{1}_{\{z\ge b\}}.
  \label{eq:paper-observed-density}
\end{equation}
Note that the parameter dependence is partly hidden in $x(z)$.

When performing the Fisher analysis, we find it more convenient to work with the unconstrained parameters
\begin{subequations}
  \label{eq:log_param}
  \begin{align}
    \eta &= \log \sigma, \\
    \alpha &= \log a.
  \end{align}
\end{subequations}
Inserting these into Eq.~\eqref{eq:paper-observed-density} and taking the logarithm, the log-likelihood for one observation is
\begin{equation}
  \ell(z; \eta, \alpha, b)
  =
  \log x(z)+\log(1+e^\eta x(z))-\alpha-\eta
  -\frac{1}{2}x(z)^2.
  \label{eq:like}
\end{equation}
This is our starting point for the Fisher information analysis.

\section{Information Bounds}

\subsection{Regular Fisher information problem}
Although $(\sigma, a, b)$ are structurally identifiable, this does not imply that $\sigma$ can be estimated with high precision. Indeed, for $\sigma \to 0$, $\log(1+\sigma X)\approx \sigma X$, and the observation model only depends on $\sigma$ through the product $a\sigma$. Likewise, for $\sigma \rightarrow \infty$, the model only depends on $\sigma$ through $a\log\sigma + b$. Using Fisher information analysis, we can quantify the information available for estimating $\sigma$ in the presence of the unknown compression law across the full range of $\sigma$ values. Through the Cram\'er--Rao bound, this information gives a lower bound on the variance of any unbiased estimator of $\sigma$. It therefore provides practical guidance for what can be hoped to learn about $\sigma$ in different speckle regimes.

To apply Fisher analysis, we first must separate the regular and irregular parts of the model. Examining Eq.~\eqref{eq:like}, $\eta$ and $\alpha$ are regular interior parameters, where perturbing either parameter perturbs the likelihood without changing its support. On the other hand, $b$ is the lower bound of the support of $Z$ as can be seen by setting $X=0$ in Eq.~\eqref{eq:model_ss_observation}. Moving $b$ changes the support itself, so the usual Fisher-based Cram\'er--Rao calculation does not apply. We apply Fisher analysis to the regular block $(\eta,\alpha)$ and treat $b$ as known. Thus, our bounds do not account for the additional uncertainty introduced by the unknown $b$ parameter. Information loss in the unknown-$b$ compression problem can only be more severe, which we empirically examine in the single-window synthetic experiments below.

The Fisher information matrix is the expected value of the negative Hessian matrix of the log likelihood, Eq.~\eqref{eq:like}, evaluated at the true parameter values. Since $\ell(z| \eta, \alpha, b)$ only depends on $z$ through $x(z)$ and $X=x(Z)$ at the true parameter values, the elements of this matrix can be written
\begin{align}
  I_{\theta\phi}
  &= \mathbb{E}_{X\sim \mathrm{Rayleigh}(1)}\left[-\frac{\partial^2\ell}{\partial \theta \partial \phi}\right] \nonumber \\
  &= \mathbb{E}_{X\sim \mathrm{Rayleigh}(1)}\left[\frac{\partial\ell}{\partial \theta}\frac{\partial\ell}{\partial \phi}\right]
  \label{eq:fish_matel}
\end{align}
where $\theta, \phi \in \{\eta, \alpha\}$, and $\ell$ is the same function as in Eq.~\eqref{eq:like}, but taken as a function of $x$ rather than $z$. The second equality in Eq.~\eqref{eq:fish_matel} follows from the equivalence between the expected curvature of the log-likelihood and the product of its first-order partial derivatives, also called \textit{score functions}.

The score functions can be computed directly from Equation \eqref{eq:like}:
\begin{align*}
    S_\eta(x)
    &\equiv \frac{\partial\ell}{\partial \eta}
    =x^2-2,
    \\
    S_\alpha(x;\sigma)
    &\equiv \frac{\partial\ell}{\partial \alpha}
    =
    -1
    -\log(1+\sigma x)
    \left(
      2+\frac{1}{\sigma x}-\frac{x}{\sigma}-x^2
    \right).
\end{align*}
Note that these expressions do not depend on the true values of $\alpha$ or $b$. Thus, the regular Fisher matrix elements, Eq.~\eqref{eq:fish_matel}, and the corresponding Cram\'er--Rao bounds are universal across models of the form Eq.~\eqref{eq:model_ss} and can only depend on $\sigma$. This is one motivation for using the log parameterization, Eq.~\eqref{eq:log_param}. Evaluating $\mathbb{E}_X\left[S_\eta(x)^2\right]$ shows that $I_{\eta\eta}=4$. $I_{\eta\alpha}$ and $I_{\alpha\alpha}$ both depend on the true value of $\sigma$ and have complicated expressions in terms of special functions. Rather than give these complex expressions, we compute these matrix elements numerically in Python using high-order direct-$r$ quadrature.

$I_{\eta\eta}=4$ is the Fisher information available about $\eta$ from a single observation of $Y$ under a known compression law. The Cram\'er--Rao inequality gives a lower bound on the variance of any unbiased estimator $\hat\eta|\alpha$ of $\eta$ given $\alpha$: $\operatorname{Var}(\hat\eta|\alpha) \ge 1/(4n)$ for a single window with $n$ independent pixels. Transforming back to the original $\sigma$ scale gives the Rayleigh limit under known compression:
\begin{equation}
  \label{eq:crb-known-compression}
  \operatorname{Var}(\hat\sigma|\alpha) \ge \sigma^2/(4n),
\end{equation}
where $\hat\sigma|\alpha$ is any unbiased estimator of $\sigma$ given $\alpha$. An unknown $\alpha$ reduces the information available about $\eta$ from a single observation. The reduced information equals the Schur complement for $\eta$:
\begin{equation*}
  I_{\eta\mid\alpha}(\sigma)
  =
  4-\frac{I_{\alpha\eta}(\sigma)^2}{I_{\alpha\alpha}(\sigma)}.
\end{equation*}
The Cram\'er--Rao inequality then gives:
\begin{equation}
  \operatorname{Var}(\hat\sigma)
  \ge
  \sigma^2 / (I_{\eta\mid\alpha}(\sigma)n),
  \label{eq:paper-crb-sigma}
\end{equation}
where $\hat\sigma$ is any unbiased estimator of $\sigma$ without conditioning on a known $\alpha$.

The parallel bound for the nuisance contrast parameter follows by profiling out $\eta$ instead:
\begin{equation*}
  I_{\alpha\mid\eta}(\sigma)
  =
  I_{\alpha\alpha}(\sigma)-\frac{I_{\alpha\eta}(\sigma)^2}{4}.
\end{equation*}
Thus, for any unbiased estimator of $a$ in the regular known-$b$ problem,
\begin{equation*}
  \operatorname{Var}(\hat a)
  \ge
  a^2 / (I_{\alpha\mid\eta}(\sigma)n).
\end{equation*}
Although $a$ is a nuisance parameter for QUS, this bound provides a useful diagnostic for the synthetic experiments below.

\begin{figure}[htbp]
  \centering
  \includegraphics[width=0.78\linewidth]{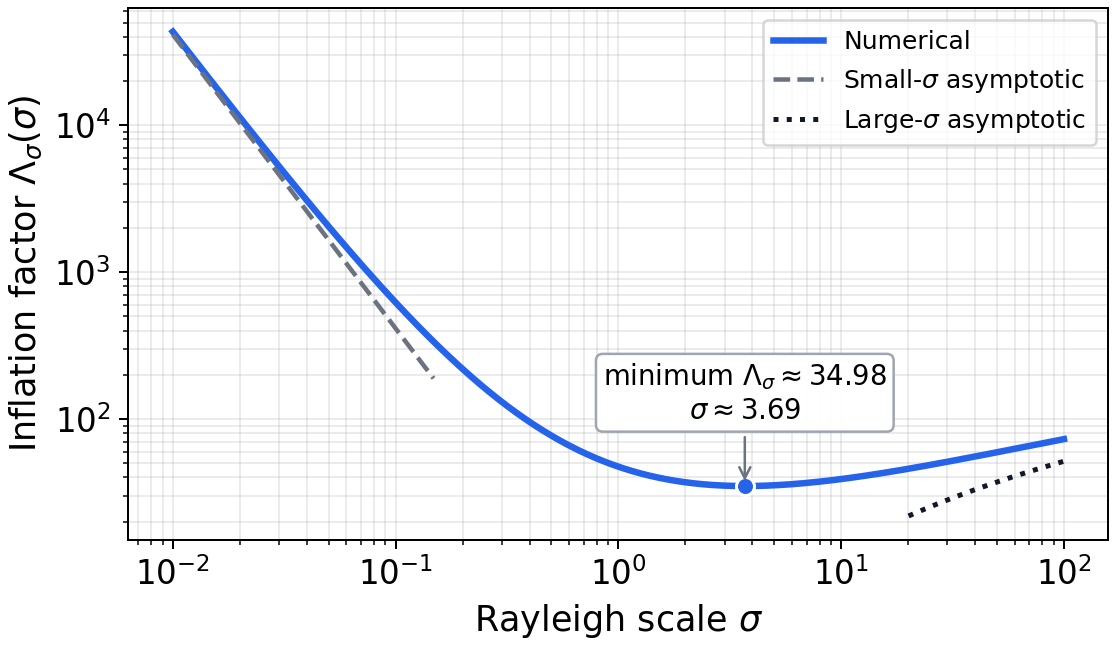}
  \caption{Single-window inflation factor $\Lambda_\sigma(\sigma)$ for the regular known-$b$ problem with unknown contrast $a$. The solid blue curve is the numerical evaluation of $4/I_{\eta\mid\alpha}(\sigma)$, and the dashed and dotted curves show the small- and large-$\sigma$ asymptotic formulas. The annotation marks the minimum inflation ratio, which is approximately $\FisherMinInflation$.}
  \label{fig:single-window-inflation}
\end{figure}

\subsection{Single-window variance inflation}

We quantify the degree to which the unknown $\alpha$ weakens estimation of $\sigma$. Taking the ratio of the lower bound in Eq.~\eqref{eq:paper-crb-sigma} to the bound in \eqref{eq:crb-known-compression}, we define the \textit{variance inflation factor}:
\begin{equation}
  \Lambda_\sigma(\sigma)
  =
  \left(1 - \frac{1}{4}\frac{I_{\alpha\eta}(\sigma)^2}{I_{\alpha\alpha}(\sigma)}\right)^{-1}
  \label{eq:paper-inflation-single}
\end{equation}
This quantity, plotted in Figure \ref{fig:single-window-inflation}, only depends on the true scale $\sigma$ and is therefore universal across the family of models described in Eq.~\eqref{eq:model_ss}. It quantifies the excess variance penalty incurred by any log compression of the form Eq.~\eqref{eq:model_ss_observation} when $a$ is unknown and $b$ is known. Numerically, $\Lambda_\sigma(\sigma)$ has a minimum of approximately $\FisherMinInflationPrecise$, attained at $\sigma \approx \FisherMinSigma$. Thus even in the most favorable regime, unknown log compression inflates the variance bound by more than an order of magnitude.

\subsection{Asymptotic behavior}

The endpoint asymptotics explain why the single-window inflation curve rises on both sides. As $\sigma \to 0$, the model behaves like $Z \approx a\sigma X+b$, so $a$ and $\sigma$ become nearly collinear and the information about $\eta$ approaches zero. On the other hand, when $\sigma \to \infty$, the model behaves like $Z \approx a\log X + a\log\sigma + b$, so $a$ and $\log\sigma$ become nearly collinear and the information about $\eta$ again approaches zero. Expanding the inflation factor, Eq.~\eqref{eq:paper-inflation-single}, in these two limits gives
\begin{align*}
    \Lambda_\sigma(\sigma)
    &=
    \frac{32}{9(4-\pi)\sigma^2} + O(\sigma^{-1}),
    \qquad \sigma \to 0,
    \\
    \Lambda_\sigma(\sigma)
    &=
    \frac{24(\log\sigma)^2}{\pi^2}+O(\log\sigma),
    \qquad \sigma \to \infty.
\end{align*}
These limiting expressions are plotted in Figure \ref{fig:single-window-inflation}.

\subsection{Multi-window pooling}

Suppose that $M$ homogeneous windows share a global contrast parameter $\alpha=\log a$ and the same known support offset $b$, but have local scanner-normalized scales $\sigma_m$, with $\eta_m=\log\sigma_m$. Each window contributes some information about the global contrast $\alpha$ which in turn reduces the excess variance penalty for estimating the local scales $\sigma_m$. We can quantify this pooling effect by applying the same Fisher analysis to the multi-window model.

If window $m$ contains $n_m$ pixels, the Fisher matrix for $(\alpha,\eta_1,\ldots,\eta_M)$ has block form
\begin{equation*}
  \mathcal I^{\mathrm{mw}}
  =
  \begin{pmatrix}
    A & c^\top \\
    c & D
  \end{pmatrix},
\end{equation*}
where
\begin{align*}
  A &= \sum_{m=1}^M n_m I_{\alpha\alpha}(\sigma_m),
  \\
  c_m &= n_m I_{\alpha\eta}(\sigma_m),
  \\
  D &= \operatorname{diag}(4n_1,\dots,4n_M).
\end{align*}
Profiling out all local scales gives the pooled information for the shared contrast parameter,
\begin{equation}
  I^{\mathrm{mw}}_{\alpha\mid\eta_{1:M}}
  =
  A-c^\top D^{-1}c
  =
  \sum_{m=1}^M n_m I_{\alpha\mid\eta}(\sigma_m),
  \label{eq:paper-schur-multi-general}
\end{equation}
showing that each window simply contributes its single-window Schur complement information $I_{\alpha\mid\eta}(\sigma_m)$ to the pooled information about $\alpha$. To obtain the local scale variances, we invert the same block matrix directly. Block inversion gives the lower-right block of the inverse information matrix,
\begin{equation*}
  \left[
    \left(\mathcal I^{\mathrm{mw}}\right)^{-1}
  \right]_{\eta\eta}
  =
  D^{-1}
  +
  D^{-1}c
  \left(I^{\mathrm{mw}}_{\alpha\mid\eta_{1:M}}\right)^{-1}
  c^\top D^{-1}.
\end{equation*}
Taking the $m$th diagonal entry and changing variables from $\eta_m$ to $\sigma_m$ gives the local scale bound
\begin{equation}
  \operatorname{Var}(\hat\sigma_m)
  \ge
  \sigma_m^2
  \left(
    \frac{1}{4n_m}
    +
    \frac{I_{\alpha\eta}(\sigma_m)^2}{16 I^{\mathrm{mw}}_{\alpha\mid\eta_{1:M}}}
  \right).
  \label{eq:paper-sigma-multi-general}
\end{equation}
The first term is the known-compression Rayleigh limit. The second term is the penalty for learning the shared $\alpha$ parameter from the full collection of windows. Note that this penalty is inversely proportional to the total information about $\alpha$, defined in Eq.~\eqref{eq:paper-schur-multi-general}, which grows with the number of windows. The same Schur complement gives the multi-window contrast bound
\begin{equation*}
  \operatorname{Var}(\hat a)
  \ge
  a^2 / I^{\mathrm{mw}}_{\alpha\mid\eta_{1:M}},
\end{equation*}
where $I^{\mathrm{mw}}_{\alpha\mid\eta_{1:M}}$ already includes the pixel counts from all windows.

\subsection{Equal-window pooling example}

To cleanly illustrate the effect of pooling, consider the case where all windows have the same size $n$ and the same scale $\sigma$. Then
\begin{equation*}
  I^{\mathrm{mw}}_{\alpha\mid\eta_{1:M}} = Mn I_{\alpha\mid\eta}(\sigma),
\end{equation*}
and the local scale bound reduces to
\begin{equation}
  \operatorname{Var}(\hat\sigma_m)
  \ge
  \sigma^2
  \left(
    \frac{1}{4n}
    +
    \frac{I_{\alpha\eta}(\sigma)^2}{16Mn I_{\alpha\mid\eta}(\sigma)}
  \right).
  \label{eq:paper-sigma-multi-equal}
\end{equation}
Relative to the known-compression bound $\sigma^2/(4n)$, the equal-window inflation factor reduces exactly to
\begin{equation}
  \Lambda_\sigma^{(M)}(\sigma)
  =
  1+\frac{\Lambda_\sigma(\sigma)-1}{M},
  \label{eq:paper-multiwindow-collapse}
\end{equation}
where $\Lambda_\sigma(\sigma)$ is the single-window inflation factor in Eq.~\eqref{eq:paper-inflation-single}. At the minimum single-window inflation $\Lambda_\sigma(\sigma) \approx \FisherMinInflationPrecise$, we require $M=\FisherBestCaseWindows$ equal windows to reduce $\Lambda_\sigma^{(M)}(\sigma)$ to at most $1.1$. This large number of windows demonstrates the practical difficulty of overcoming the unknown-compression penalty, even in the most favorable scenarios.

\section{Estimation and Evaluation}
\label{sec:estimation}
We validate our theoretical findings by comparing the Cram\'er--Rao bounds to empirical estimator variance in synthetic and real-data experiments. Here, we discuss the estimation procedure and evaluation methods used in our experiments. For $M$ independent homogeneous windows, with $n_m$ pixels in window $m$, shared compression parameters $(a,b)$, and local scales $\sigma_m$, define $\boldsymbol{\sigma}=(\sigma_1,\ldots,\sigma_M)$.
On the common support $z_{mi}\ge b$ for all observations, the full log-likelihood is
\begin{equation}
  \begin{aligned}
    \ell(a,b,\boldsymbol{\sigma})
    &=
    \sum_{m=1}^{M}
    \sum_{i=1}^{n_m}
    \bigg[
      \log x_{mi}
      + \log\!\left(1+\sigma_m x_{mi}\right)
      \\
      &\qquad\qquad
      - \log a
      - \log\sigma_m
      - \frac{1}{2}x_{mi}^2
    \bigg],
  \end{aligned}
  \label{eq:empirical-full-log-likelihood}
\end{equation}
where $x_{mi}=x(z_{mi};\sigma_m,a,b)$ is the inverse image from Eq.~\eqref{eq:paper-inverse-image} evaluated with the local scale $\sigma_m$. Equation \eqref{eq:empirical-full-log-likelihood} includes the single-window case by setting $M=1$ and $n_1=n$. To match our theoretical setting, our experiments focus on the regular known-$b$ problem. However, we also include unknown-$b$ fits to empirically demonstrate the additional instability caused by the nonregular support edge.

For any fixed $(a,b)$, maximizing Eq.~\eqref{eq:empirical-full-log-likelihood} over each local scale gives the asymptotically unbiased estimate
\begin{equation}
  \label{eq:scale-estimator}
  \hat\sigma_m(a,b)
  =
  \left[
    \frac{1}{2n_m}
    \sum_{i=1}^{n_m}
    \left\{
      \exp\!\left(\frac{z_{mi}-b}{a}\right)-1
    \right\}^2
  \right]^{1/2}.
\end{equation}
We use this relation to profile the local scales $\hat\sigma_m$ out of Eq.~\eqref{eq:empirical-full-log-likelihood}. To match the assumptions of the theoretical Cram\'er--Rao bounds derived above, we insert the known $b$ value and optimize only over $\alpha=\log a$ for numerical stability. We find the optimal $\alpha$ using a bounded golden-section search.

In our results, we also consider the unknown-$b$ case as an empirical diagnostic. Because $b$ is a support parameter, we use the support-safe parameterization
\begin{equation*}
  \alpha=\log a,
  \qquad
  \delta=\log(z_{\min}-b),
  \qquad
  b=z_{\min}-e^\delta,
\end{equation*}
where $z_{\min}$ is the minimum observed pixel value among the fitted windows. This guarantees that the fitted support edge remains below the observed data. We initialize $(a,b)$ with the Seabra--Sanches moment and boundary estimates, then optimize the profiled log-likelihood jointly over $(\alpha,\delta)$ using L-BFGS with bounded search ranges. Since the Cram\'er--Rao bounds apply to unbiased estimators, the finite-sample comparisons below should be read primarily as variance checks in regimes where empirical bias is small; the relative-RMSE axes in the figures use the same caveat.

\subsection{Theory--experiment comparison}

We use dimensionless error measures to compare the experiments with the theoretical bounds across scales and window sizes. When the population parameters are known, as in the synthetic experiments, we report the scaled variances
\begin{equation}
  V_m=\frac{n_m\operatorname{Var}(\hat\sigma_m)}{\sigma_m^2},
  \qquad
  V_a=\frac{n_m\operatorname{Var}(\hat a)}{a^2}.
  \label{eq:comparison-scaled-variances}
\end{equation}
The known-compression Rayleigh limit is $V_m\ge 1/4$; for a single window with unknown contrast, $V_m\ge 1/I_{\eta\mid\alpha}(\sigma_m)=\Lambda_\sigma(\sigma_m)/4$ and $V_a\ge 1/I_{\alpha\mid\eta}(\sigma_m)$. For pooled windows, the scale prediction is obtained by multiplying Eq.~\eqref{eq:paper-sigma-multi-general} by $n_m/\sigma_m^2$.

For the real-data experiments, the population scale is unknown. From the RF-envelope samples $x_{mi}$ in window $m$, we form the reference estimate $\sigma_m^{\mathrm{RF}}=(2n_m)^{-1/2}(\sum_i x_{mi}^2)^{1/2}$. We then measure scale recovery using
\begin{equation}
  E_m
  =
  \frac{n_m(\hat\sigma_m-\sigma_m^{\mathrm{RF}})^2}
       {(\sigma_m^{\mathrm{RF}})^2}.
  \label{eq:comparison-real-data-error}
\end{equation}
The average of $E_m$ over windows is the normalized mean squared error shown in the real-data results. Because the same RF samples determine $\sigma_m^{\mathrm{RF}}$ and the log-compressed pixels used to estimate $\hat\sigma_m$, the two scale estimates have correlated sampling error. Under Eq.~\eqref{eq:model_ss}, independent Rayleigh pixels, and negligible estimator bias, the matched-sample prediction therefore retains only the excess uncertainty from estimating the compression contrast:
\begin{subequations}
  \label{eq:comparison-matched-sample}
  \begin{align}
    \mathbb{E}[E_m]
    &\simeq
    \frac{1}{I_{\eta\mid\alpha}(\sigma_m^{\mathrm{RF}})}-\frac{1}{4},
    && M=1,
    \label{eq:comparison-matched-sample-single}
    \\
    \mathbb{E}[E_m]
    &\simeq
    \frac{n_m I_{\alpha\eta}(\sigma_m^{\mathrm{RF}})^2}
         {16I^{\mathrm{mw}}_{\alpha\mid\eta_{1:M}}},
    && M>1.
    \label{eq:comparison-matched-sample-pooled}
  \end{align}
\end{subequations}
The pooled expression is the second term in Eq.~\eqref{eq:paper-sigma-multi-general} after normalization, and all Fisher terms are evaluated at the RF reference scales. As a contrast-estimation diagnostic, we similarly use $E_{a,m}=n_m(\hat a-a)^2/a^2$, whose single-window prediction is $\mathbb{E}[E_{a,m}]\simeq 1/I_{\alpha\mid\eta}(\sigma_m^{\mathrm{RF}})$.

\section{Experiments}
\label{sec:experiments}

\subsection{Synthetic validation}

We first use synthetic estimation experiments to test our theoretical bounds in a setting where our model assumptions are strictly satisfied. We generate single- or multi-window data using Eq.~\eqref{eq:model_ss} and compare the empirical scaled variances $V_m$ and $V_a$, defined in Eq.~\eqref{eq:comparison-scaled-variances} of the theory--experiment comparison above, with the corresponding Cram\'er--Rao bounds. Throughout, we quantify the uncertainty in these empirical metrics by drawing $2000$ bootstrap replicates of the underlying simulation runs and reporting pointwise $95\%$ intervals in each figure.

\paragraph{Single-window.} We first simulate independent homogeneous windows from Eq.~\eqref{eq:model_ss} with one window per replicate. The compression parameters are fixed at $a=20/\log(10)$ and $b=0$. The window size is $32\times 32$ pixels ($n=1024$), and the scanner-normalized Rayleigh scale is swept over
$\sigma\in\{10^{-2},10^{-1},0.3,1,3.5,10,31.6,10^2\}$ with $200$ replicates per scale. For each replicate, we fit the regular known-$b$ profile likelihood described above by optimizing only $\alpha=\log a$ with a bounded golden-section search over $\alpha\in[\log 0.5,\log 200]$ and an optimization budget of $80$ steps. We then plug the fitted $\hat a$ and the known $b$ into the scale estimator, Eq.~\eqref{eq:scale-estimator}, to obtain $\hat\sigma$. We also directly estimate $\sigma$ from the simulated envelope samples as the known-compression baseline. Finally, we fit the profiled unknown-$(a,b)$ model with an L-BFGS iteration budget of $80$ steps to demonstrate the added instability when $b$ is unknown, which is not captured by our regular Fisher information analysis.

\begin{figure}[t]
  \centering
  \includegraphics[width=0.97\linewidth]{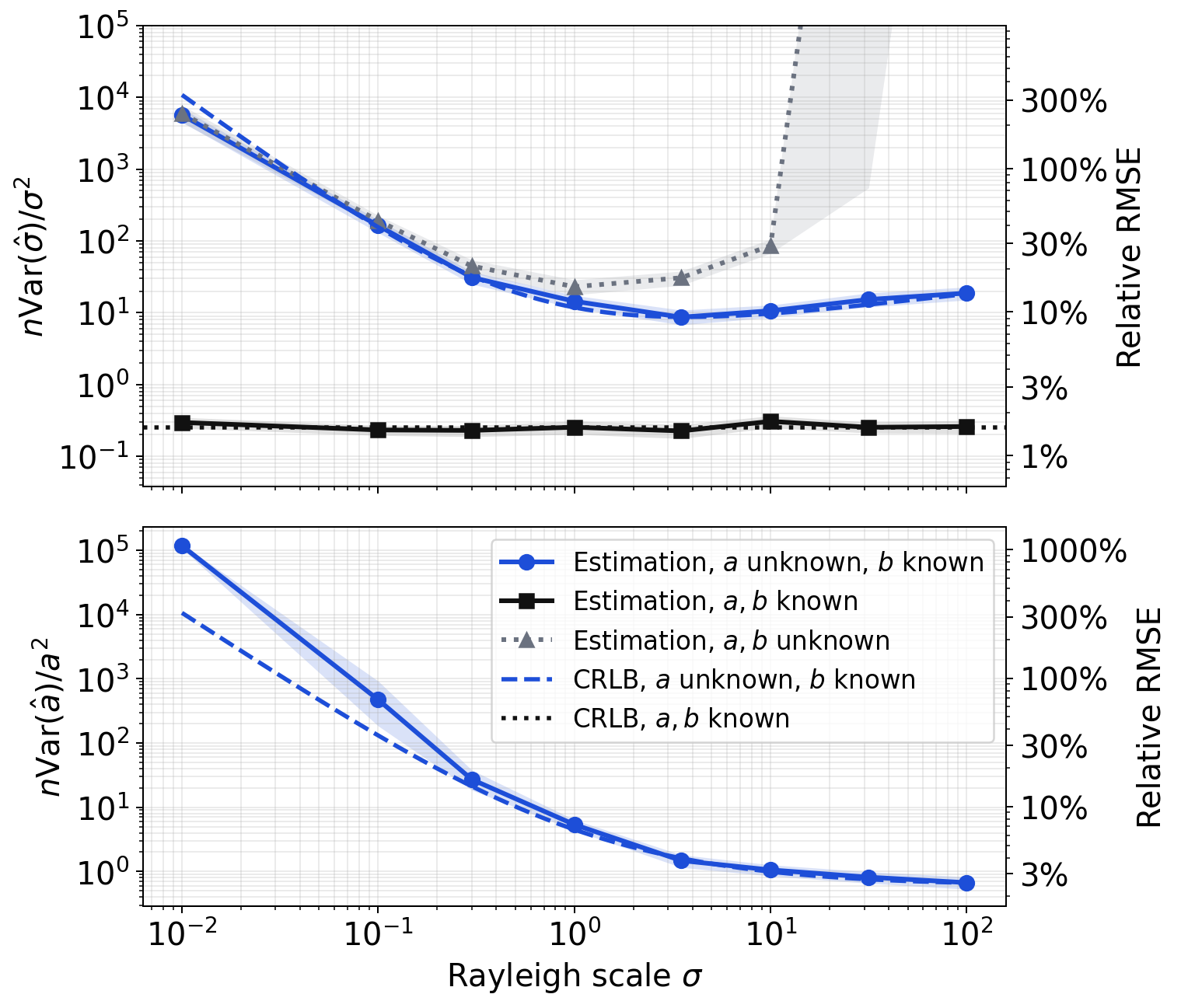}
  \caption{Single-window synthetic validation. The top and bottom y-axes approximate $V_m$ and $V_a$, respectively. The top panel shows scaled variance for $\hat\sigma$, and the bottom panel shows scaled variance for $\hat a$. The right axes convert scaled variance to relative RMSE for the $32\times32$ windows when estimator bias is negligible. The gray dotted curve in the top panel is the measured variance in the case of unknown $a$ and $b$. The shaded bands are $95\%$ bootstrap intervals based on 2000 replicates of the 200 simulation runs for each point.}
  \label{fig:single-window-validation}
\end{figure}

For this experiment, the theory--experiment comparison gives $V_m\ge 1/I_{\eta\mid\alpha}(\sigma)=\Lambda_\sigma(\sigma)/4$ (see Eqs.~\eqref{eq:comparison-scaled-variances}, \eqref{eq:paper-crb-sigma}, and \eqref{eq:paper-inflation-single}). Figure~\ref{fig:single-window-validation} shows that the known-$b$, unknown-$a$ estimator closely follows this predicted bound for $V_m$ over most of the tested scale range. The $200$ replicates per scale are sufficient for the main variance trends except in the most poorly identified regimes, exemplified by the disagreement with the empirical variance in $\hat a$ for $\sigma < 0.3$. The right axes show relative RMSE assuming that the estimator bias is negligible. The single-window estimator reaches roughly the $10\%$ relative-error scale over the middle of the range, but the bound prevents it from improving much below that level for a $32\times32$ window.

The unknown-$(a,b)$ fits clearly demonstrate the added instability from the irregular boundary parameter, especially at large $\sigma$ values. This is expected since, in the large-sigma limit, the model is exactly collinear in $a\log\sigma$ and $b$.

\paragraph{Multi-window.} We next repeat the regular known-$b$ estimation problem with $M$ equal-sized homogeneous windows sharing the same contrast parameter and the same true local Rayleigh scale. This matches the equal-window pooling example described above: the data are generated with a common $\sigma$, but the likelihood still treats the $M$ local scales as separate parameters $\sigma_m$. Here $a=20/\log(10)$, $b=0$, each window has size $16\times 16$ pixels ($n=256$), and $\sigma$ is swept over $\{10^{-2},10^{-1},1,10,10^2\}$. For each value of $\sigma$ and
$M\in\{2^k:k=0,\ldots,10\}$, we generate $100$ independent pooled datasets, estimate the shared $a$ with the same known-$b$ profile likelihood and an optimization budget of $80$ steps, and then compute $\hat\sigma_m$ for each of the $M$ windows using Eq.~\eqref{eq:scale-estimator}. The empirical metric $V_m$ from the theory--experiment comparison is computed over the resulting $100M$ local scale estimates for each $(\sigma,M)$ setting. This experiment checks the equal-window variance bound in Eq.~\eqref{eq:paper-sigma-multi-equal} and the pooled inflation factor in Eq.~\eqref{eq:paper-multiwindow-collapse}: pooling should reduce only the excess nuisance penalty, while the known-compression Rayleigh term remains.

\begin{figure}[t]
  \centering
  \includegraphics[width=0.98\linewidth]{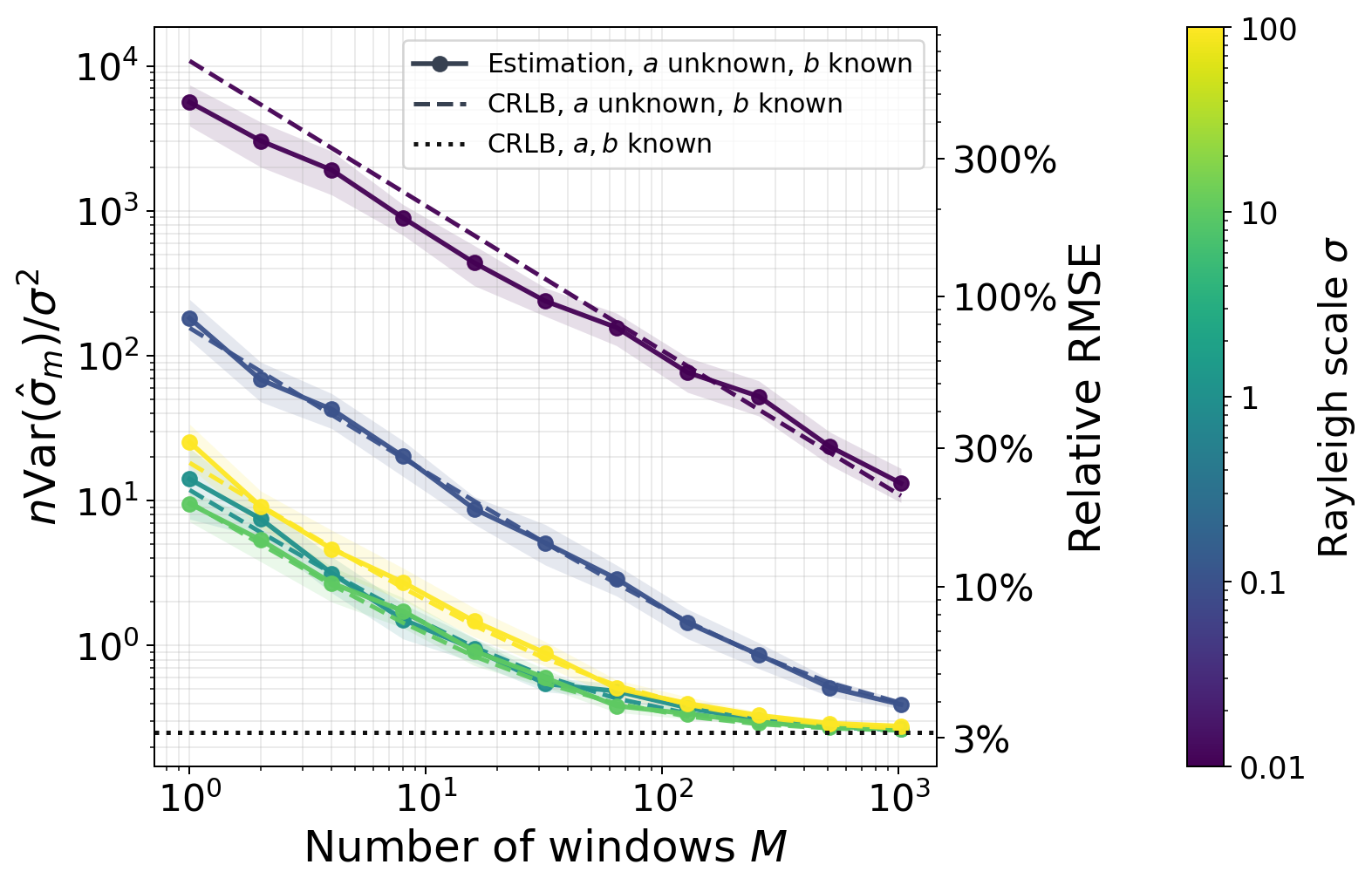}
  \caption{Pooled equal-window synthetic validation. The y-axis approximates $V_m$. The x-axis is the number of pooled windows, and color encodes the true Rayleigh scale $\sigma$. Each colored pair compares empirical scaled variance with its corresponding equal-window bound. The horizontal reference is the known-compression limit. The right axis converts scaled variance to relative RMSE for the $16\times16$ windows. The shaded bands are $95\%$ bootstrap intervals based on 2000 replicates of the 100 simulation runs for each point.}
  \label{fig:pooling-by-m}
\end{figure}

Figure~\ref{fig:pooling-by-m} confirms this scaling empirically across the tested scale values. As $M$ increases, the measured $V_m$ follows the normalized equal-window variance bound from the theory--experiment comparison, shown by the dashed curves, and approaches the known-compression floor. Thus pooling removes the excess penalty from the shared unknown contrast but does not beat the Rayleigh limit. The relative-RMSE axis gives the corresponding error scale for the chosen $16\times16$ window size. For this window size, the known compression limit has roughly 3\% relative RMSE. The smaller values of $\sigma$ require more than $10^3$ windows to approach this level of relative error, suggesting that it may not be feasible to overcome the unknown-compression penalty in some practical scenarios.

Together, the single-window and multi-window synthetic experiments demonstrate both aspects of the theoretical result. Unknown compression produces the predicted single-window penalty, while pooling reduces the excess nuisance contribution at the predicted $1/M$ rate towards the known-compression Rayleigh floor. However, the large number of windows required to approach the Rayleigh floor raises the question of whether it is possible to obtain precise scale estimates for real ultrasound data.

\subsection{Real-data validation}

\paragraph{The OASBUD dataset.} The OASBUD contains 100 breast-lesion records (52 malignant and 48 benign) from 78 women, with two orthogonal RF scans per lesion (200 scans total). For this experiment, we use only the first stored scan from each lesion record, yielding 100 analyzed scans, one per lesion \cite{piotrzkowska-wroblewska_open_2017}. Each scan provides beamformed radio-frequency (RF) ultrasonic echoes. Access to the RF data lets us form log-compressed images with known display parameters, and compare B-mode estimates against RF-derived reference scales from the same measured envelope data. We note that this experimental procedure makes use of the beamformed RF data for the selection of Rayleigh-like windows. This information would not be available in an application context, where only the B-mode image is accessible. The goal of this experiment is not to show the direct clinical utility of our estimation procedure. Rather, we evaluate the extent to which the theoretical bounds that we derived above describe the variance pattern when estimating the Rayleigh scale from real B-mode images in the presence of unknown log compression.

We treat each beamformed RF pixel as an individual observation. Each of the scans that we analyze contains $510$ RF echo lines along the $\qty{38}{\milli\meter}$ transducer aperture, each of which is sampled at $\qty{40}{\mega\hertz}$ to produce evenly spaced samples every $\qty{0.0192}{\milli\meter}$. The number of samples per echo line varies with the depth of the scan, from $1040$ samples for a $\qty{2}{\centi\meter}$ depth to $2864$ samples for a $\qty{6}{\centi\meter}$ depth. Due to specular scattering, clustering, and other effects, the RF envelope amplitudes badly violate the Rayleigh assumption, in general. We proceed by selecting Rayleigh-like windows from the RF scans as described next.

\begin{figure}[htbp]
  \centering
  \includegraphics[width=1\linewidth]{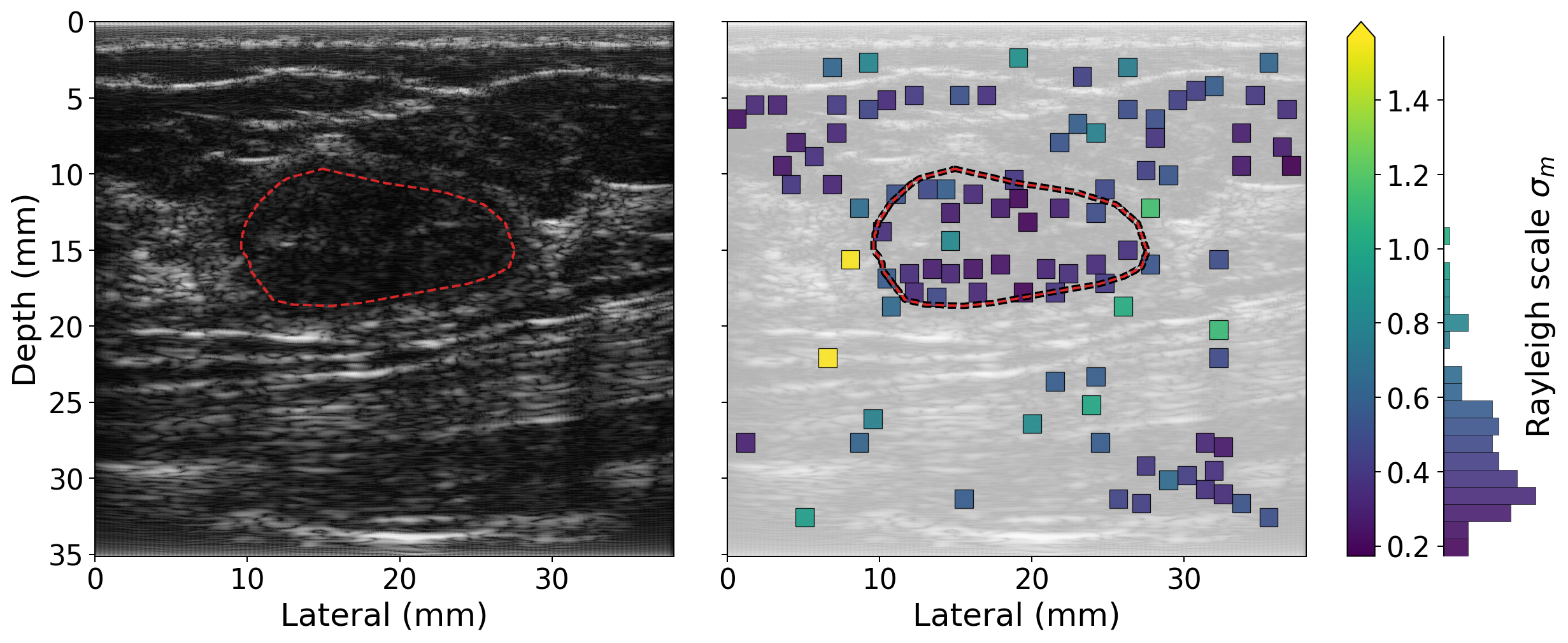}
  \caption{Window selection for OASBUD scans. The left panel is a B-mode image generated from the RF scan showing a benign lesion (BI-RADS category 3) outlined in red. The right panel overlays the selected windows colored according to their RF-derived reference scale $\sigma_m^{\mathrm{RF}}$. The histogram plots the distribution of $\sigma_m^{\mathrm{RF}}$ across the selected windows.}
  \label{fig:selected-window-example}
\end{figure}

\begin{figure}[htbp]
  \centering
  \includegraphics[width=1\linewidth]{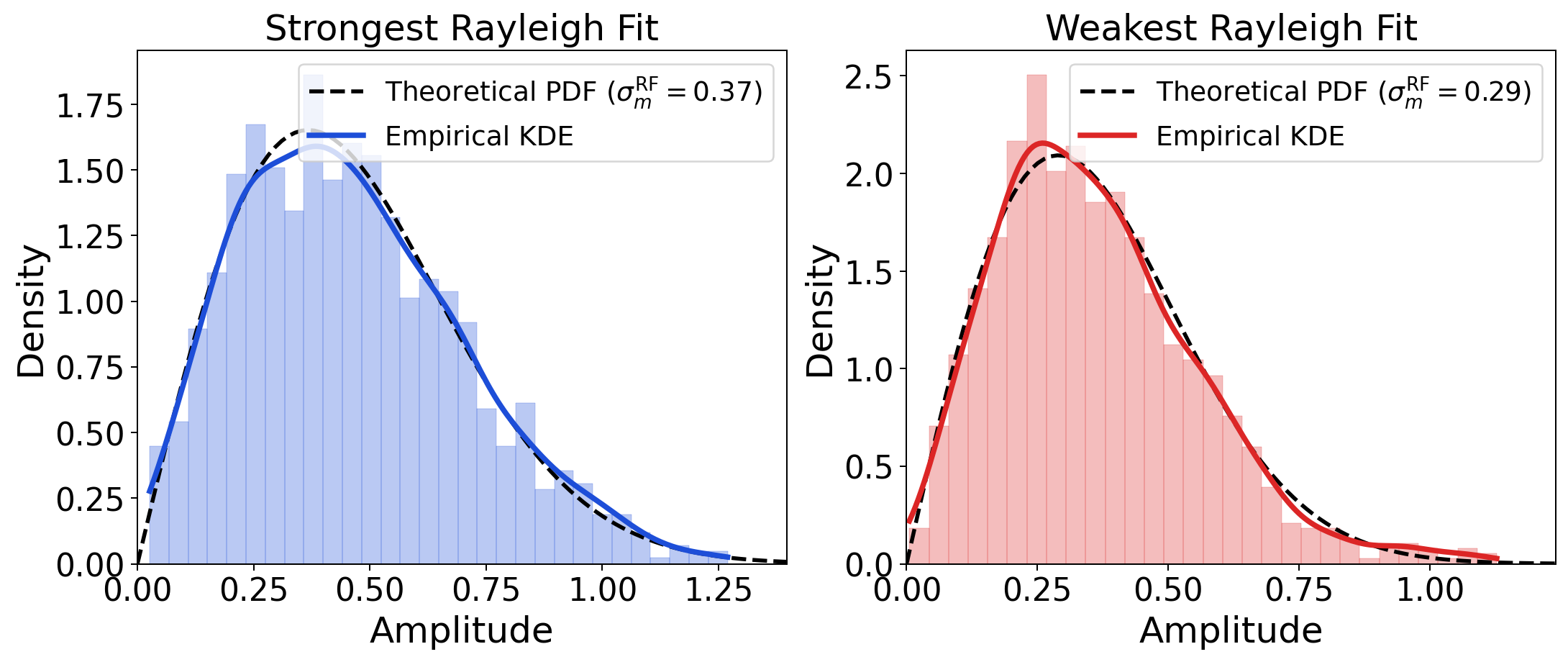}
  \caption{Empirical versus theoretical distributions for best and worst windows. The left panel shows the window with the strongest Rayleigh agreement (lowest Anderson--Darling statistic). The right panel shows the weakest agreement. Both show good agreement between the empirical kernel density estimate (KDE) and the theoretical Rayleigh density with scale $\sigma_m^{\mathrm{RF}}$.}
  \label{fig:selected-window-histogram}
\end{figure}

\paragraph{Window selection.} To select Rayleigh-like windows, we first discard the upper and lower $5\%$ of the data from each scan to remove shallow and deep edge artifacts. We then define a grid of overlapping candidate windows over the RF grid. Each window is $16$ echo lines wide and uses $64$ samples along the axial direction. This corresponds to a physical $\qty{1.19}{\milli\meter} \times \qty{1.23}{\milli\meter}$ rectangle. We let adjacent candidate windows overlap by $12$ echo lines laterally and $48$ samples axially. For each candidate window with envelope samples $x_i$, we compute the normalized squared amplitudes $u_i=x_i^2/[2(\sigma_m^{\mathrm{RF}})^2]$. Under the Rayleigh model, these values should be distributed as $\mathrm{Exp}(1)$, so we rank the candidate windows by the Anderson--Darling statistic for this exponential fit. Finally, we greedily select the top $100$ non-overlapping candidate windows. Figure~\ref{fig:selected-window-example} shows the selected windows overlaid on the B-mode image of the first scan in the dataset. Figure~\ref{fig:selected-window-histogram} compares the empirical and theoretical distributions for two of the selected windows: the one with the strongest agreement with the Rayleigh model and the one with the weakest agreement. Even in the latter case, the histogram of the RF-envelope amplitudes still closely follows the Rayleigh density.

We use the selected windows to test parameter recovery. For each RF envelope window, we form a B-mode image window using compression with $a=20/\log 10$ and $b=0$. We then fit the known-$b$, unknown-$a$ estimator to a single window or pooled windows.

\begin{figure}[tbp]
  \centering
  \includegraphics[width=1\linewidth]{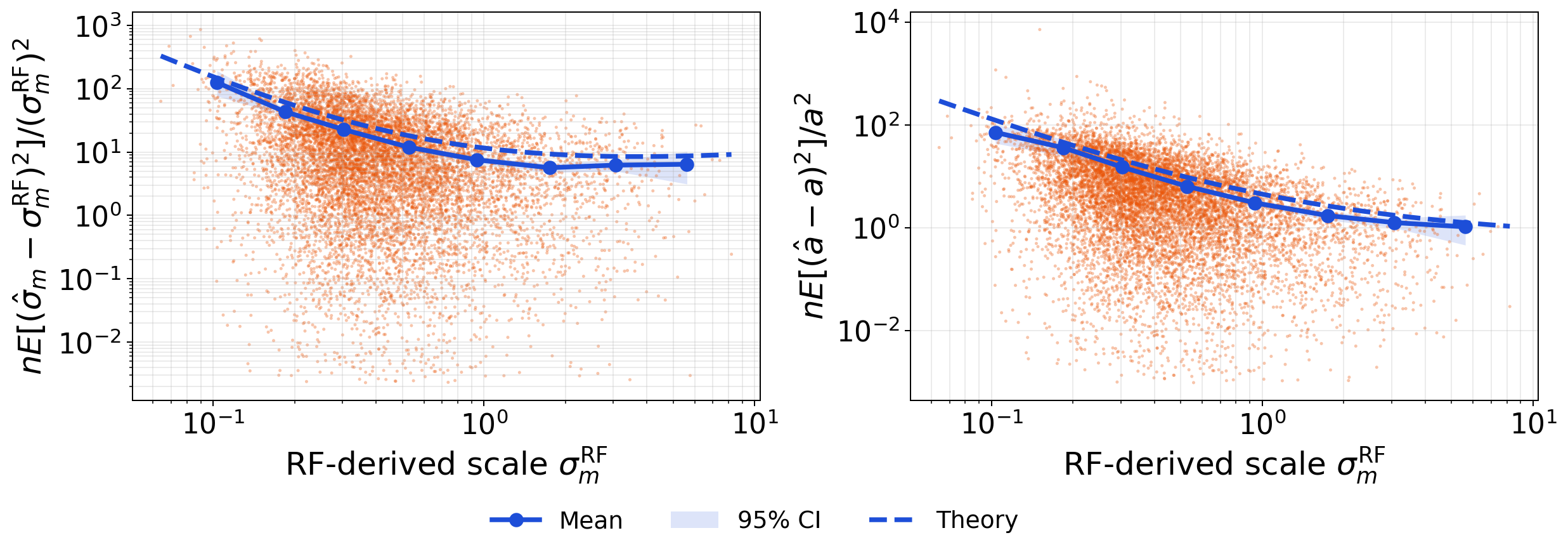}
  \caption{Single-window OASBUD results. The left and right y-axes approximate $E_m$ and $E_{a,m}$, respectively. Orange points show individual squared errors from 100 simulation runs. Solid blue points show the MSE binned by $\sigma_m^{\mathrm{RF}}$. Shaded bands are the $95\%$ bootstrap intervals based on 2000 replicates of the 100 simulation runs. Dashed blue curves show the theoretical predictions under matched conditions.}
  \label{fig:oasbud-single-validation}
\end{figure}

\paragraph{Single-window.} We fit each selected window separately and group the normalized squared recovery errors from Eq.~\eqref{eq:comparison-real-data-error} into eight logarithmically spaced bins of $\sigma_m^{\mathrm{RF}}$. Within each bin, we average errors within scan and then weight the contributing scans equally. Pointwise $95\%$ intervals come from $2000$ bootstrap resamples of scans, retaining all selected windows from each sampled scan. As shown in Figure~\ref{fig:oasbud-single-validation}, the resulting error curve closely follows the shape of the matched-sample prediction in Eq.~\eqref{eq:comparison-matched-sample-single} but lies systematically below it, suggesting model misspecification in the real-data windows.

\paragraph{Multi-window.} Our multi-window experiments test how well the pooled matched-sample prediction in Eq.~\eqref{eq:comparison-matched-sample-pooled} describes the observed error levels as the number of real-data windows increases. For each OASBUD scan in our analysis, we form $50$ random permutations of the $100$ selected Rayleigh-like windows. For each permutation, we fit the pooled estimator using the first $M\in\{1,2,3,5,10,20,30,50\}$ windows. Each fit returns a shared contrast parameter estimate $\hat a$ and the window-specific scale estimates $\hat\sigma_m$. The window sets are nested to reduce shot noise among the fits within a permutation. At $M=100$, all $100$ selected windows are used in the fit, so we avoid refitting across the permutations. We assign each selected window to one of three fixed quantile bins of $\sigma_m^{\mathrm{RF}}$; its bin stays fixed across nested $M$ values and permutations. For each $M$ and bin, we average the window errors within each scan across permutations, then average the scan-level means equally across scans. The pooled intervals use the same scan-bootstrap draws as the single-window analysis.

\begin{figure}[tbp]
  \centering
  \includegraphics[width=0.7\linewidth]{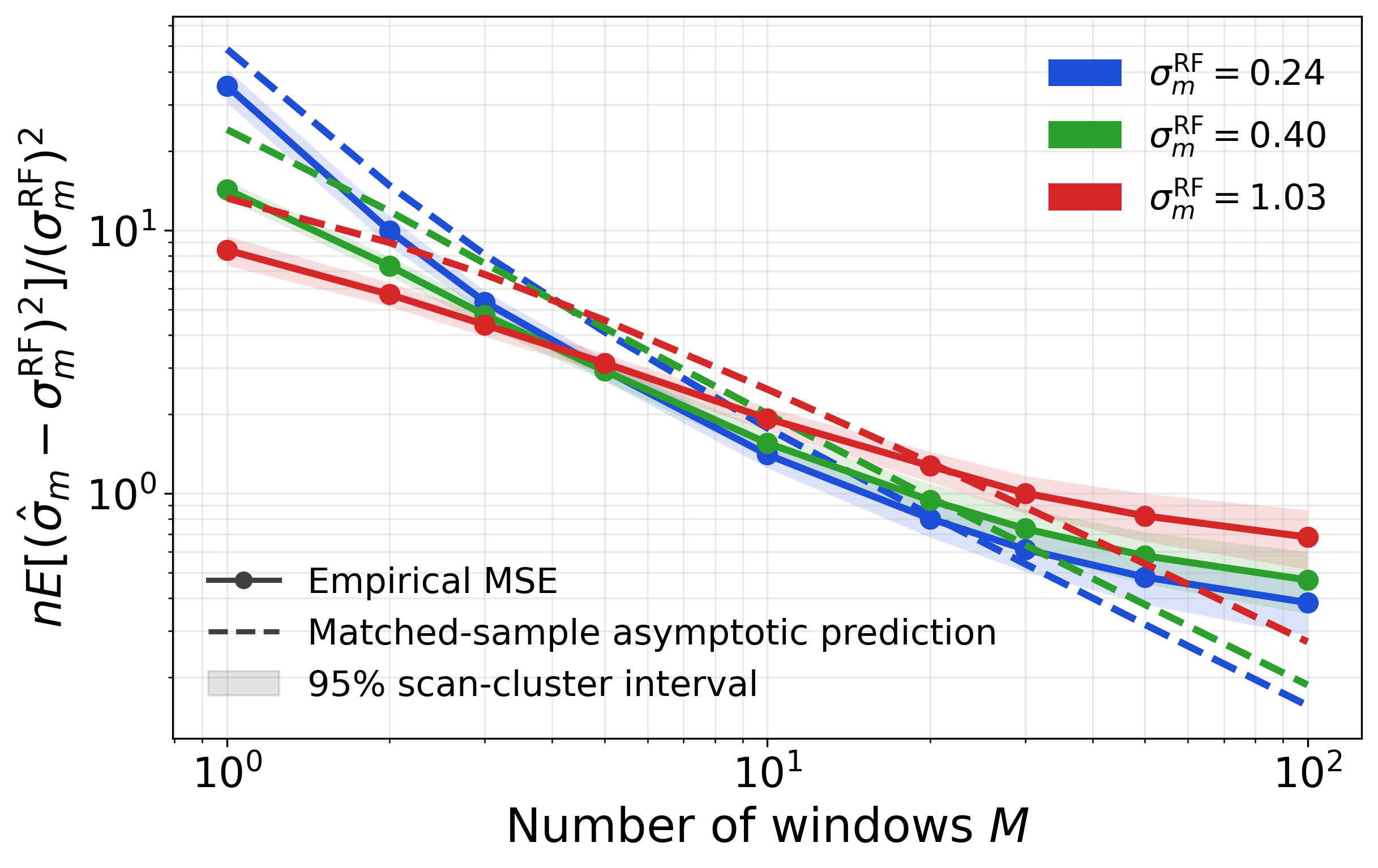}
  \caption{Pooled OASBUD results. The y-axis approximates $E_m$. Solid curves show MSE in Rayleigh scale estimates versus pooling size $M$ based on 100 simulations. Shaded bands are the $95\%$ bootstrap intervals based on 2000 replicates of the 100 simulation runs. Dashed curves show the theoretical predictions under matched conditions.}
  \label{fig:oasbud-pooled-validation}
\end{figure}

As shown in Figure \ref{fig:oasbud-pooled-validation}, the theoretical curves correctly predict the shape of the pooled error curves for $M<10$, the curve crossings near $M=4$, and the sorting of the measured errors by scale across the full range of $M$. However, for $M>10$, the shape of the curves deviates from theory and appears to saturate. This suggests an asymptotic bias in the pooled estimates which may arise from model misspecification or from selection bias due to our window-selection procedure. Nevertheless, the clear signature of the theoretical pooling effect in the $M<10$ regime shows how the unknown-compression penalty can be partially mitigated by pooling multiple windows, even in real B-mode images.

\section{Discussion}

We establish two constraints on Rayleigh-scale estimation from B-mode images. First, the physical Rayleigh scale $\sigma^*$ is not identifiable without knowledge of the scanner-dependent factor $b/c$ in Eq.~\eqref{eq:identified-parameters-sigma}. Log-compressed data instead identify the scanner-normalized scale $\sigma$. This hard identifiability constraint persists regardless of the number of available pixels or image windows. However, when display settings are shared across windows, ratios of Rayleigh scales remain identifiable even though each absolute physical scale is not. Comparisons across scans acquired with different scanners or display settings therefore require knowledge of the compression parameters or external calibration. Second, even $\sigma$, which is technically identifiable, can be inefficient to estimate when the contrast parameter $a$ must be learned from the same observations.

Our theoretical analysis characterizes the loss of precision due to the unknown contrast parameter $a$, which arises not just because of the log transformation. When the compression law is known, the transformation can be inverted, and the Fisher information about the Rayleigh scale agrees with the information available from the RF envelope samples. However, when $a$ is unknown, B-mode signals strongly couple information about $\sigma$ and $a$ as expressed in the off-diagonal elements of the Fisher information matrix. In the known-$b$ problem, this coupling inflates the single-window Cram\'er--Rao variance bound by at least approximately $\FisherMinInflation\times$ relative to known compression. Equivalently, the corresponding standard-deviation bound is inflated by approximately $5.9\times$, even at the most favorable value of $\sigma$.

Synthetic experiments corroborate our theoretical findings. In the single-window experiment, the known-$b$, unknown-$a$ estimator follows the predicted scale-variance bound over most of the tested range. The weaker agreement at small and large $\sigma$ reflects the two approximate collinearities identified by the asymptotic analysis. The unknown-$(a,b)$ fits show instability at large $\sigma$, highlighting that the regular known-$b$ result should not be interpreted as a complete model of uncertainty when the full display law is unknown. In the multi-window experiment, the empirical scale variances follow the predicted pooling curves and approach the known-compression Rayleigh floor. Thus, for sufficiently large number of windows, pooling can reduce the excess uncertainty from jointly estimating $a$, addressing the efficiency issue highlighted by the single-window analysis. However, in our synthetic experiments, the number of windows required to approach the Rayleigh limit is large, suggesting that it may remain difficult to overcome the unknown-compression penalty in practice.

We tested the applicability of our theoretical results to real B-mode images using the OASBUD dataset. Our single-window results show clear agreement between the measured estimator error and error predicted by our information-theoretic analysis but with a systematic downward bias in the empirical error. In the context of multiple pooled windows with a shared contrast parameter, the theory has a clear correspondence with the measured error for $M<10$ but begins to deviate substantially for larger $M$. This suggests model specification issues or biases introduced by our window selection procedure. In the pooled results, the theory describes the initial decrease in error, the curve crossings near $M=4$, and the ordering by scale, but the empirical curves begin to saturate for $M>10$. These deviations do not contradict the Cram\'er--Rao calculation, which assumes the exact Rayleigh model and unbiased estimation. Departures from the Rayleigh assumption, spatial dependencies among pixels, and the RF-based window-selection procedure may all contribute to the discrepancy. The OASBUD experiment therefore supports the qualitative signature of the unknown-compression penalty and the benefit of pooling, but it also highlights the practical challenges of applying the theory to real B-mode images.

\section{Limitations and Future Work}

Our analysis uses the Rayleigh model, which only takes into account homogeneous, fully developed speckle. Real tissue can contain coherent components, sparse scatterers, attenuation variation, shadowing, and mixtures of tissue types. A future direction is to extend our analysis to richer models, like the Rician, Nakagami, and homodyned-K models, to clarify whether the same compression penalty appears for other QUS targets \cite{destrempes2025theoretical}. Such extensions will also require careful treatment of finite-window estimation error, since recent homodyned-K studies have shown that parameter accuracy and uncertainty depend strongly on sample size and sample correlation \cite{tehrani2024homodyned}. The relevance of these extensions is supported by recent clinical RF studies in which Nakagami and nonparametric backscatter statistics detected early hepatic steatosis, although their values were also influenced by fibrosis and other tissue-level covariates \cite{lin2024clinical}.

The display model is also quite simplified. We analyze finite-offset log compression with unknown contrast and, in the Fisher calculation, known support offset. Clinical systems usually also apply time-gain compensation, clipping, quantization, scan conversion, persistence, despeckling, and other vendor-specific processing \cite{li2022reproducibility}. Experimental robustness studies further show that the resulting B-mode texture depends on beamforming and display dynamic-range choices \cite{seoni2023texture}. The unknown-$b$ experiments already suggest that support-related display parameters add instability beyond the regular known-$b$ bound.

Finally, our bounds assume independent samples, whereas neighboring ultrasound samples are spatially correlated. Consequently, the number of samples in a window may overstate the amount of independent information available, and extensions of the theory should account explicitly for spatial correlation or effective sample size \cite{tehrani2024homodyned}. In addition, the OASBUD experiment is a controlled recovery test in which B-mode images are generated from available RF data, rather than a direct evaluation on archived clinical B-mode images. Future work should therefore validate the theory on vendor-produced B-mode images or calibrated phantom data and address the nonregular unknown-$b$ problem.

\section{Conclusion}

Unknown log compression substantially reduces the precision of Rayleigh-scale estimation from B-mode data. In the finite-offset log-Rayleigh model, only the scanner-normalized scale is identifiable, and the regular known-$b$ single-window variance bound is inflated by at least $\FisherMinInflation$ when the contrast parameter is unknown. Pooling windows with shared display settings reduces this excess penalty at the predicted $1/M$ rate, but it does not remove the local Rayleigh sampling term. B-mode recovery is therefore possible, but the resulting scale estimates are scanner-normalized and carry a large nuisance-parameter uncertainty unless compression is known or well pooled.

\bibliographystyle{SageV}
\bibliography{references}

\begin{thebibliography}{10}
\providecommand{\url}[1]{\texttt{#1}}
\providecommand{\urlprefix}{URL }
\expandafter\ifx\csname urlstyle\endcsname\relax
  \providecommand{\doi}[1]{DOI:\discretionary{}{}{}#1}\else
  \providecommand{\doi}{DOI:\discretionary{}{}{}\begingroup
  \urlstyle{rm}\Url}\fi
\providecommand{\eprint}[2][]{\url{#2}}

\bibitem{li2022reproducibility}
Li MD, Cheng MQ, Chen LD et~al.
\newblock Reproducibility of radiomics features from ultrasound images:
  influence of image acquisition and processing.
\newblock \emph{European radiology} 2022; 32(9): 5843--5851.

\bibitem{seoni2023texture}
Seoni S, Matrone G and Meiburger KM.
\newblock Texture analysis of ultrasound images obtained with different
  beamforming techniques and dynamic ranges--a robustness study.
\newblock \emph{Ultrasonics} 2023; 131: 106940.

\bibitem{goodman1975speckle}
Goodman JW.
\newblock Statistical properties of laser speckle patterns.
\newblock In Dainty JC (ed.) \emph{Laser Speckle and Related Phenomena}.
  Berlin: Springer, 1975.
\newblock pp. 9--75.
\newblock \doi{10.1007/BFb0111436}.

\bibitem{burckhardt1978speckle}
Burckhardt CB.
\newblock Speckle in ultrasound {B}-mode scans.
\newblock \emph{IEEE Transactions on Sonics and Ultrasonics} 1978; 25(1): 1--6.
\newblock \doi{10.1109/T-SU.1978.30978}.

\bibitem{wagner1987statistical}
Wagner RF, Insana MF and Brown DG.
\newblock Statistical properties of radio-frequency and envelope-detected
  signals with applications to medical ultrasound.
\newblock \emph{Journal of the Optical Society of America A} 1987; 4(5):
  910--922.
\newblock \doi{10.1364/JOSAA.4.000910}.

\bibitem{destrempes2025theoretical}
Destrempes F and Cloutier G.
\newblock Theoretical foundations of the echo envelope statistical modeling: A
  tutorial.
\newblock \emph{IEEE Transactions on Ultrasonics, Ferroelectrics, and Frequency
  Control} 2025; 72(12): 1566--1581.
\newblock \doi{10.1109/TUFFC.2025.3632084}.

\bibitem{christensen2024systematized}
Christensen AM, Rosado-Mendez IM and Hall TJ.
\newblock A systematized review of quantitative ultrasound based on first-order
  speckle statistics.
\newblock \emph{IEEE transactions on ultrasonics, ferroelectrics, and frequency
  control} 2024; 71(7): 872--886.

\bibitem{cloutier2021primer}
Cloutier G, Destrempes F, Yu F et~al.
\newblock Quantitative ultrasound imaging of soft biological tissues: A primer
  for radiologists and medical physicists.
\newblock \emph{Insights into Imaging} 2021; 12: 127.
\newblock \doi{10.1186/s13244-021-01071-w}.

\bibitem{jia2022radiomics}
Jia Y, Yang J, Zhu Y et~al.
\newblock Ultrasound-based radiomics: Current status, challenges and future
  opportunities.
\newblock \emph{Medical Ultrasonography} 2022; 24(4): 451--460.
\newblock \doi{10.11152/mu-3248}.

\bibitem{li2021steatosisdl}
Li B, Tai DI, Yan K et~al.
\newblock Accurate and generalizable quantitative scoring of liver steatosis
  from ultrasound images via scalable deep learning.
\newblock \emph{World Journal of Gastroenterology} 2022; 28(22): 2494.

\bibitem{crawford1993compensation}
Crawford DC, Bell DS and Bamber JC.
\newblock Compensation for the signal processing characteristics of ultrasound
  {B}-mode scanners in adaptive speckle reduction.
\newblock \emph{Ultrasound in Medicine and Biology} 1993; 19(6): 469--485.
\newblock \doi{10.1016/0301-5629(93)90123-6}.

\bibitem{kaplan1994logcompressed}
Kaplan D and Ma Q.
\newblock On the statistical characteristics of log-compressed {Rayleigh}
  signals: Theoretical formulation and experimental results.
\newblock \emph{Journal of the Acoustical Society of America} 1994; 95(3):
  1396--1400.

\bibitem{dutt1996logcompressed}
Dutt V and Greenleaf JF.
\newblock Adaptive speckle reduction filter for log-compressed {B}-scan images.
\newblock \emph{IEEE Transactions on Medical Imaging} 1996; 15(6): 802--813.
\newblock \doi{10.1109/42.544498}.

\bibitem{prager2003decompression}
Prager RW, Gee AH, Treece GM et~al.
\newblock Decompression and speckle detection for ultrasound images using the
  homodyned {K}-distribution.
\newblock \emph{Pattern Recognition Letters} 2003; 24(4--5): 705--713.
\newblock \doi{10.1016/S0167-8655(02)00176-9}.

\bibitem{sanches2003compensation}
Sanches JM and Marques JS.
\newblock Compensation of log-compressed images for 3-{D} ultrasound.
\newblock \emph{Ultrasound in Medicine and Biology} 2003; 29(2): 239--253.
\newblock \doi{10.1016/S0301-5629(02)00710-X}.

\bibitem{vegas2011realistic}
Vegas-S{\'a}nchez-Ferrero G, Mart{\'i}n-Mart{\'i}nez D, Casaseca-de-la Higuera
  P et~al.
\newblock Realistic log-compressed law for ultrasound image recovery.
\newblock In \emph{18th IEEE International Conference on Image Processing}. pp.
  2029--2032.
\newblock \doi{10.1109/ICIP.2011.6115877}.

\bibitem{payen2013echopower}
Payen T, Coron A, Lamuraglia M et~al.
\newblock Echo-power estimation from log-compressed video data in dynamic
  contrast-enhanced ultrasound imaging.
\newblock \emph{Ultrasound in Medicine and Biology} 2013; 39(10): 1826--1837.
\newblock \doi{10.1016/j.ultrasmedbio.2013.03.022}.

\bibitem{seabra2008modeling}
Seabra J and Sanches JM.
\newblock Modeling log-compressed ultrasound images for radio frequency signal
  recovery.
\newblock In \emph{30th Annual International Conference of the IEEE Engineering
  in Medicine and Biology Society}. pp. 426--429.
\newblock \doi{10.1109/IEMBS.2008.4649181}.

\bibitem{seabra2012rf}
Seabra J and Sanches JM.
\newblock {RF} ultrasound estimation from {B}-mode images.
\newblock In Sanches JM, Laine AF and Suri JS (eds.) \emph{Ultrasound Imaging:
  Advances and Applications}. Boston, MA: Springer, 2012.
\newblock pp. 3--24.
\newblock \doi{10.1007/978-1-4614-1180-2_1}.

\bibitem{piotrzkowska-wroblewska_open_2017}
{Piotrzkowska-Wr{\'o}blewska} H, {Dobruch-Sobczak} K, Byra M et~al.
\newblock Open access database of raw ultrasonic signals acquired from
  malignant and benign breast lesions.
\newblock \emph{Medical Physics} 2017; 44(11): 6105--6109.
\newblock \doi{10.1002/mp.12538}.

\bibitem{tehrani2024homodyned}
Tehrani AK, Cloutier G, Tang A et~al.
\newblock Homodyned k-distribution parameter estimation in quantitative
  ultrasound: Autoencoder and bayesian neural network approaches.
\newblock \emph{IEEE Transactions on Ultrasonics, Ferroelectrics, and Frequency
  Control} 2024; 71(3): 354--365.

\bibitem{lin2024clinical}
Lin CH, Ho MC, Lee PC et~al.
\newblock Clinical performance of ultrasonic backscatter parametric and
  nonparametric statistics in detecting early hepatic steatosis.
\newblock \emph{Ultrasonics} 2024; 142: 107391.

\end{thebibliography}

\end{document}